\documentclass{iaus}
\usepackage{graphicx}
\usepackage{natbib}

\title[S266.~~Membership and CMD Decontamination] %% give here short title %%
{Membership Probability via Control Field \\ Colour-Magnitude Decontamination}

\author[W.\,J.\,B. Corradi, F.\,F.\,S. Maia \& J.\,F.\,C. Santos Jr.] %% give here short author list %%
{Wagner\,J.\,B.\,Corradi$^1$, Francisco\,F.\,S.\,Maia$^2$ \and Jo\~ao\,F.\,C.\,Santos\,Jr.$^3$}
\affiliation{Universidade Federal de Minas Gerais, ICEx - Dept. de F\'\i sica \\ 
CP 702, 30123-970 - Av. Ant\^onio Carlos, 6627, Belo Horizonte, Brazil \\
email: $^1${\tt wag@fisica.ufmg.br}, \ $^2${\tt ffsmaia@ufmg.br}, \ $^3${\tt jsantos@fisica.ufmg.br}}

\pubyear{2009}
\volume{266}  %% insert here IAU Symposium No.
\pagerange{119--126}
\jname{Star Clusters - Galactic Building Blocks throughout Time and Space}
\editors{Richard de Grijs \& Jacques R.~D. L\'epine, eds.}
\begin{document}

\maketitle

\begin{abstract}
The open clusters' fundamental physical parameters are important tools to 
understand the formation and evolution of the Galactic disk and as grounding tests 
for star formation and evolution models. However only a small fraction of the
known open clusters in the Milky Way has precise determination of distance, 
reddening, age, metallicity, radial velocity and proper motion. One of the major 
problems in determining these parameters lies on the difficulty to separate cluster 
members from field stars and to assign membership. We propose a decontamination 
method by employing 2MASS data in the encircling region of  
the clusters NGC1981, NGC2516, NGC6494 and M11. We present a decontaminated 
CMD of these objects showing the membership probabilities and structural 
parameters as derived from King profile fitting. 
 
\end{abstract}

\firstsection % if your document starts with a section,
\section*{Introduction}

Despite the large number of known open clusters many of them have been 
discovered recently and have not been studied in detail or do not have their 
parameters determined \citep[DAML02 -][]{DAML02,Mermilliod,Kharchenko2005}. 
Sky surveys like 2MASS \citep{2mass} produced large amounts of near-IR data 
and have contributed to the discovery of even more objects.

The determination of the fundamental parameters of open clusters provide important 
tools to the investigation of the Galactic disk and star formation models. However, 
contamination by field-stars and strong reddening severely hamper the detection and
characterization of clusters, specially towards the Galactic plane or the Galactic 
center.

We devised a method to remove the field-star contamination on colour-magnitude 
diagrams (CMD) by sampling the population from a nearby control field and then 
removing it from the cluster's CMD. The 2MASS catalog was selected owing to its 
whole sky coverage and freedom to extract data from spatially unlimited regions. 
Also, near-IR wavelengths are particularly sensitive to discriminate cluster stars 
from the contaminating field for young stellar systems \citep[e.g.,][]{sbb05}.

\section*{Data}

Vizier ({\it http://vizier.u-strasbg.fr/viz-bin/VizieR}) was used to extract
near-infrared photometric data from 2MASS  in circular fields centered on objects 
NGC\,1981, NGC\,2516, NGC\,6494 and M\,11; with the clusters' center coordinates 
taken from DAML02. The data comply with the 2MASS Level 1 Requirement 
(namely $J < 15.8$, $H < 15.1$ $K < 14.3$ mag.) and was extracted within 30 arcmin 
of the cluster's center coordinates. 

\begin{figure}[htb]
\centering
\includegraphics[scale=0.28]{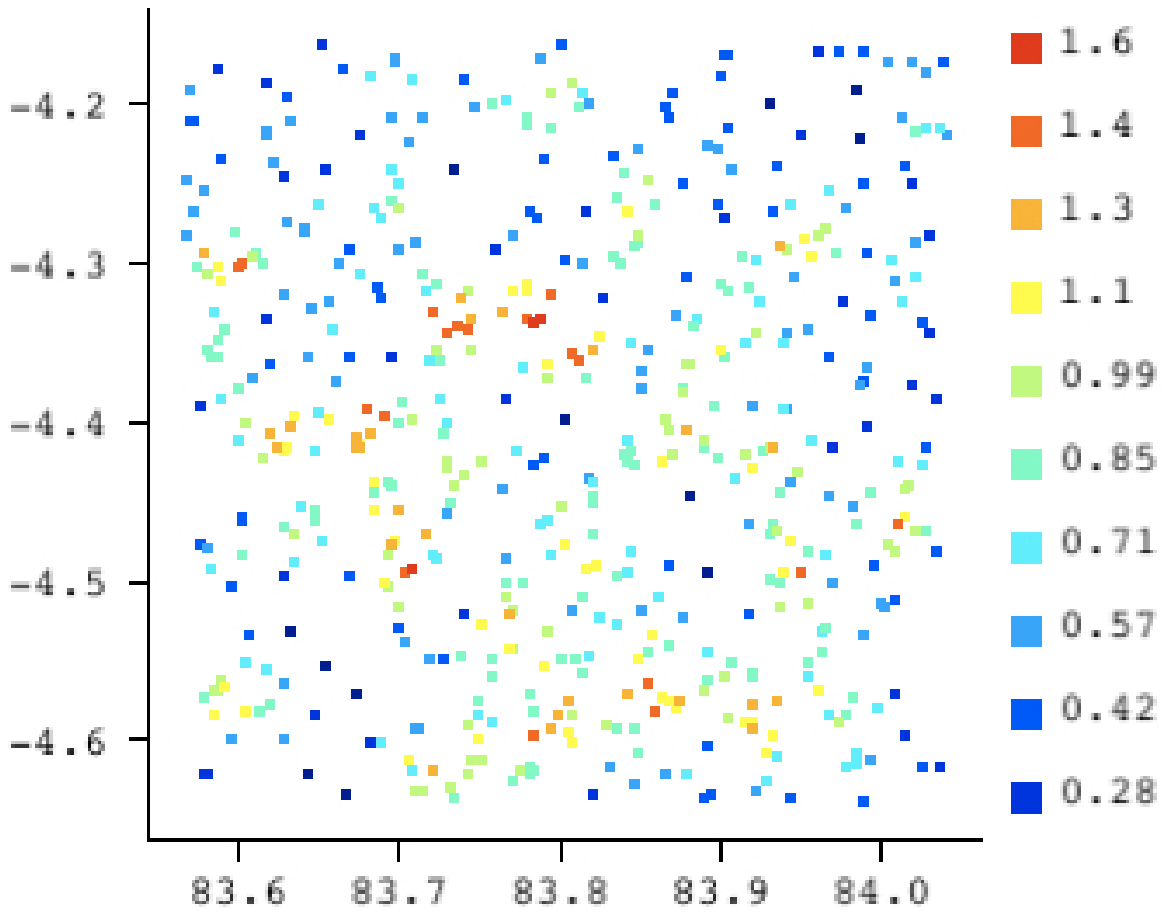}\includegraphics[scale=0.28]{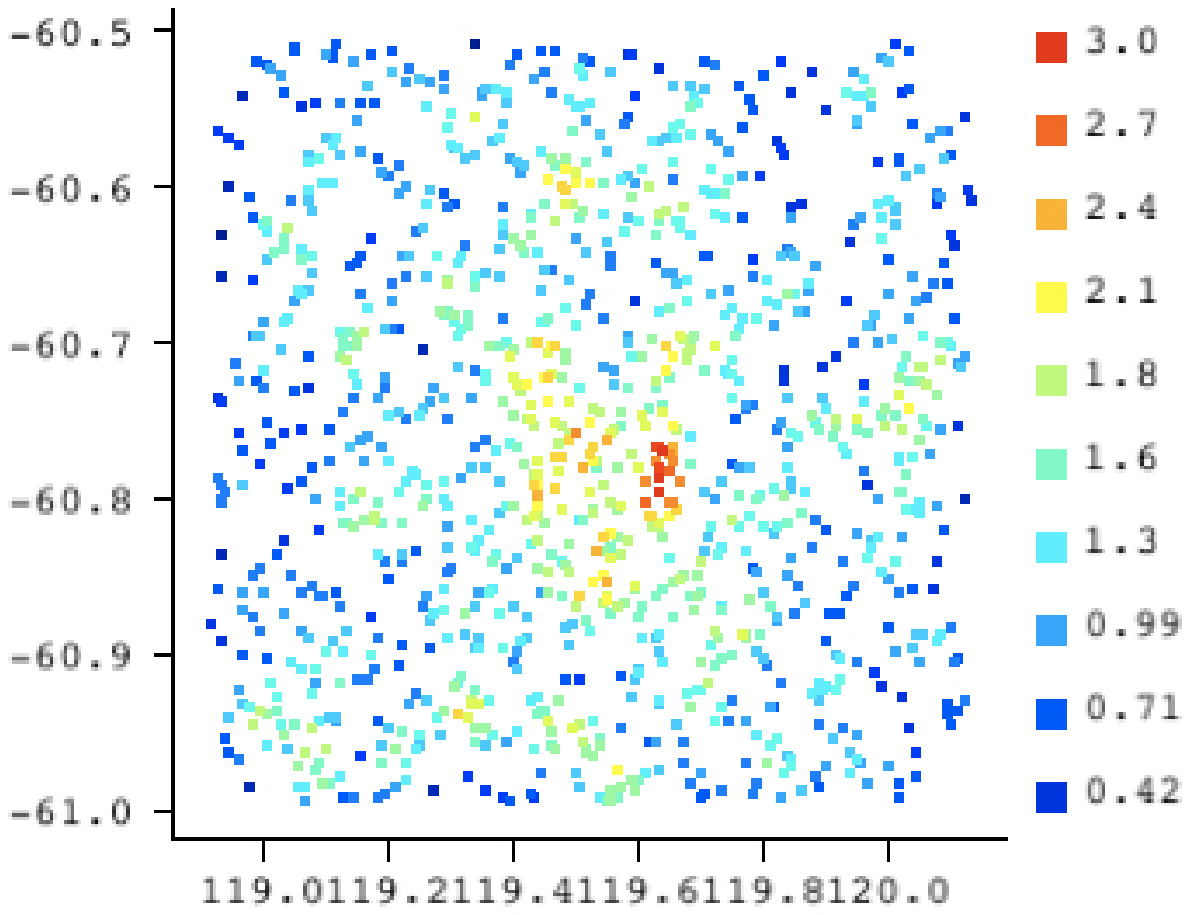}\includegraphics[scale=0.28]{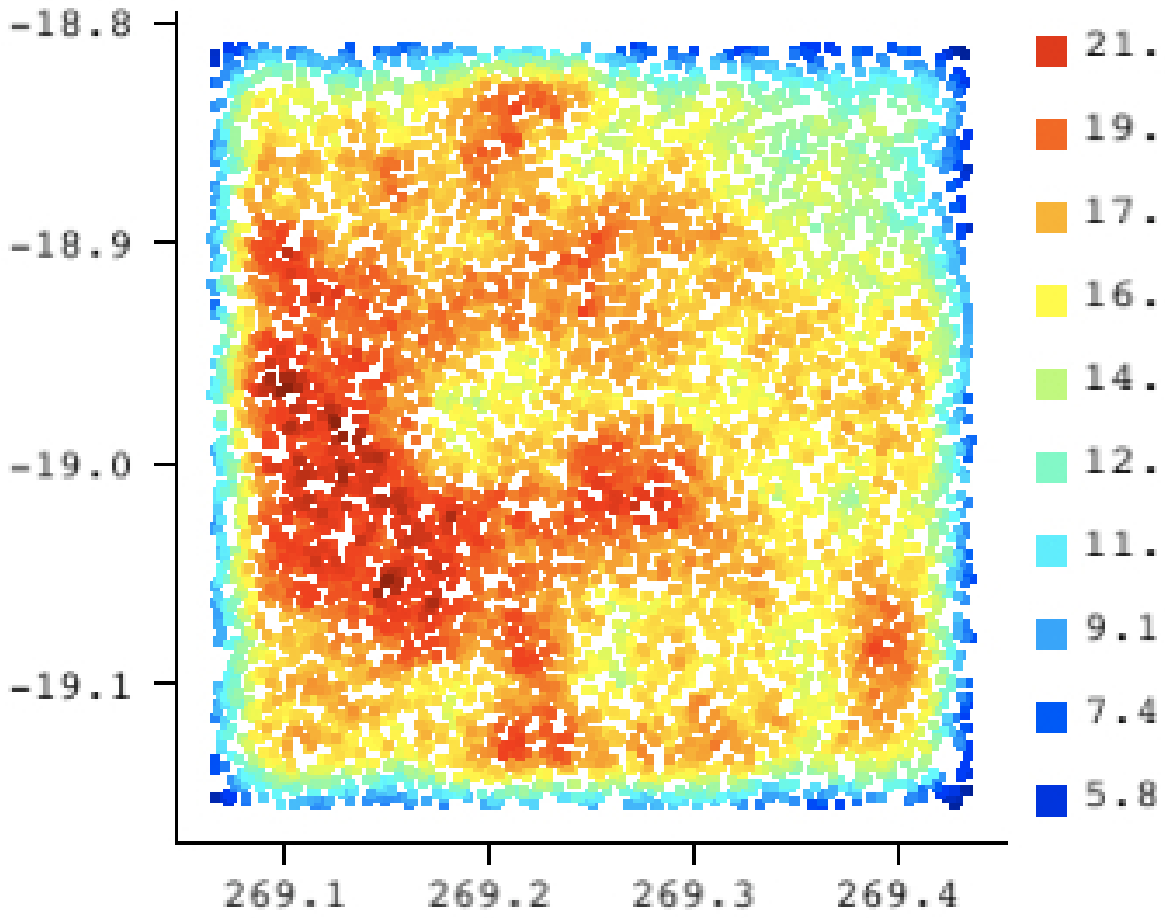}\includegraphics[scale=0.28]{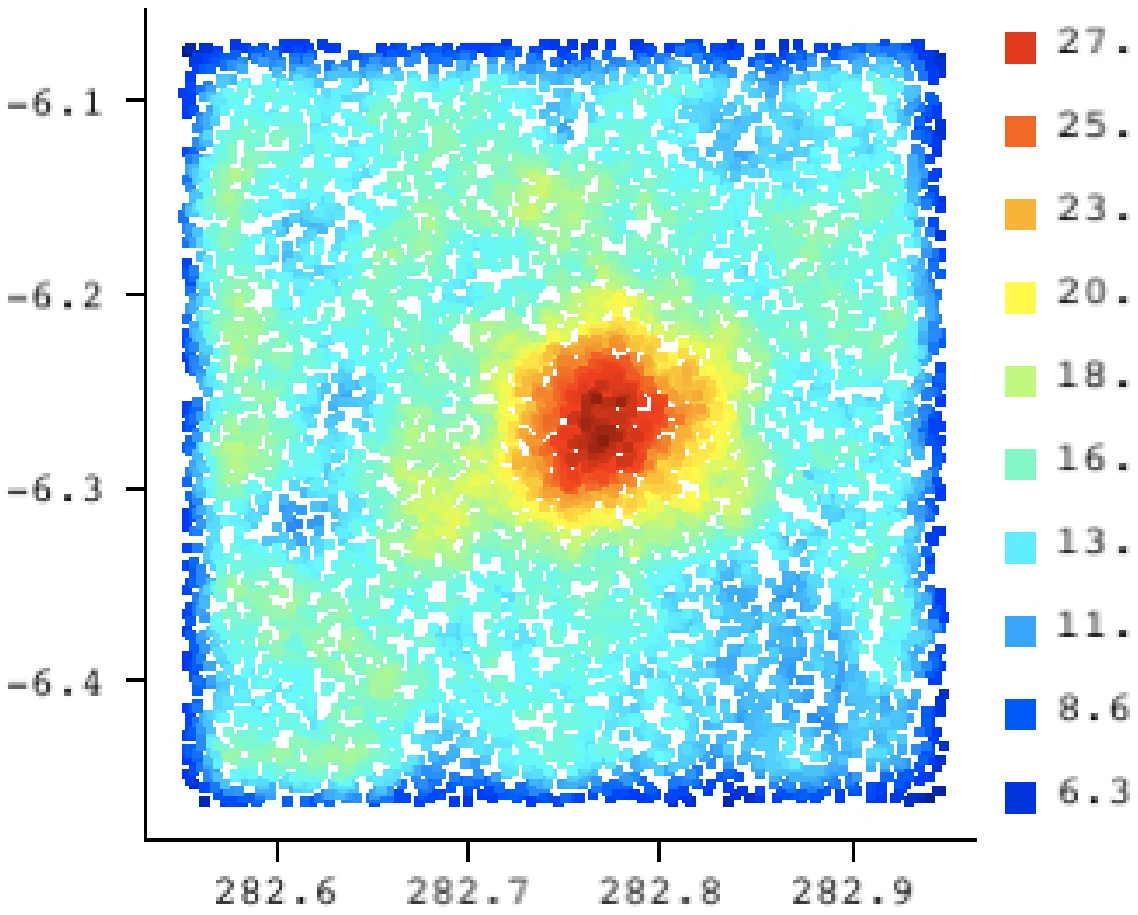}
\includegraphics[scale=0.27]{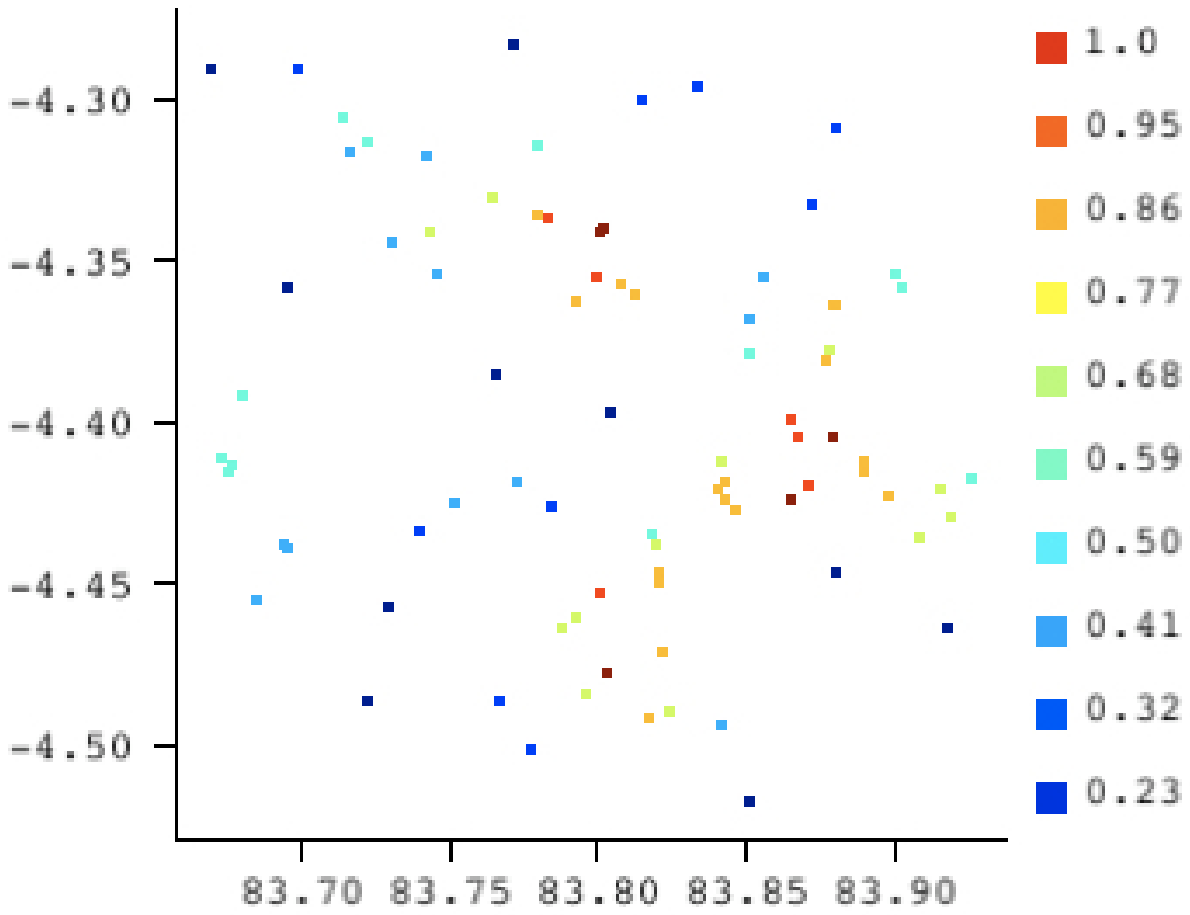}\includegraphics[scale=0.27]{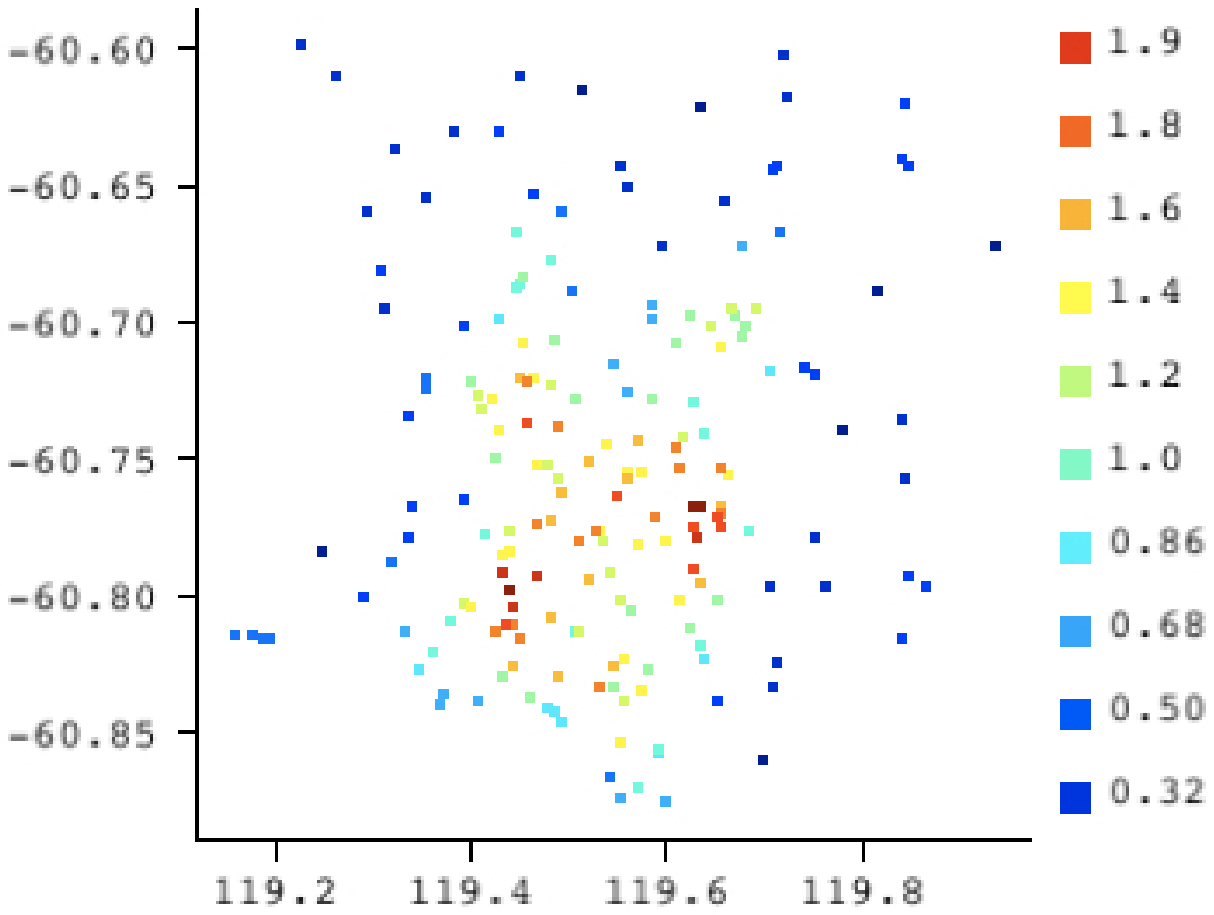}\includegraphics[scale=0.27]{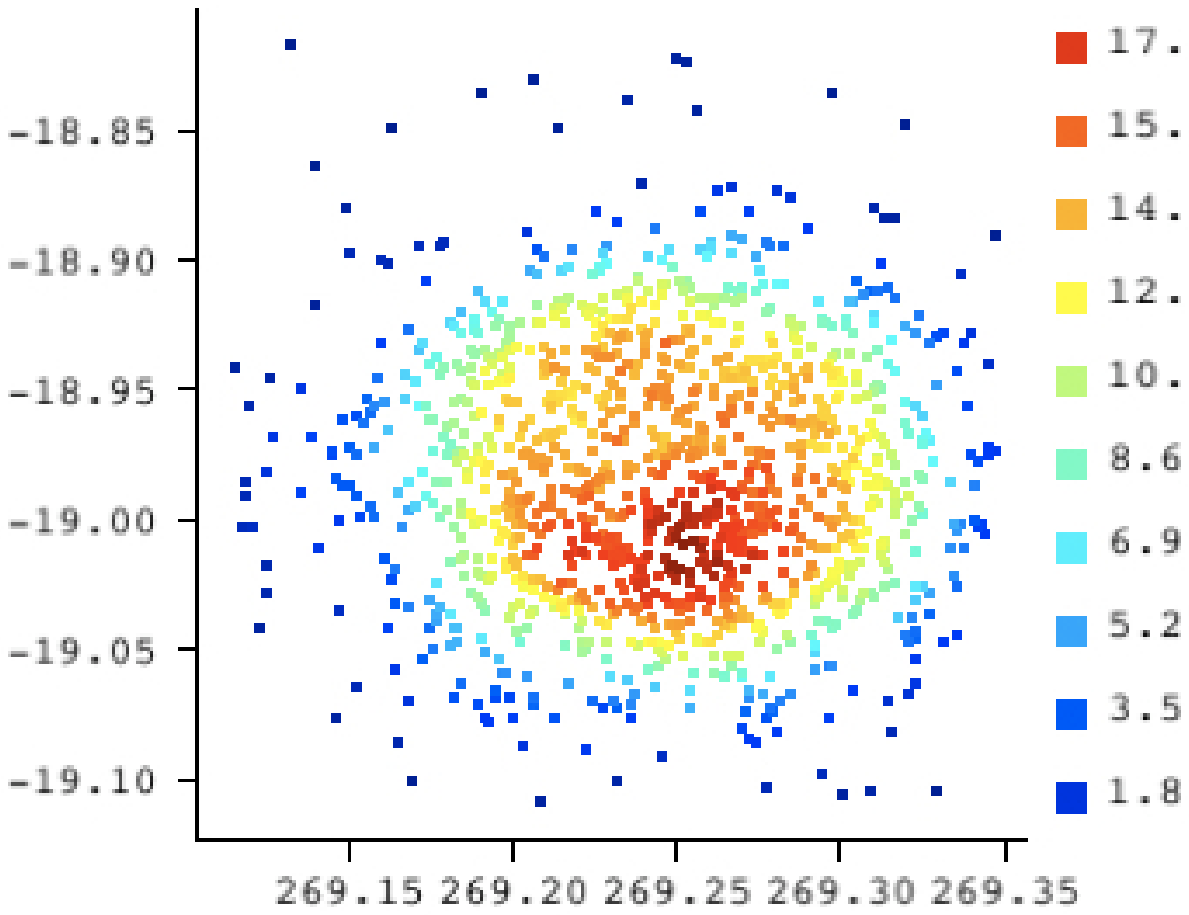}\includegraphics[scale=0.27]{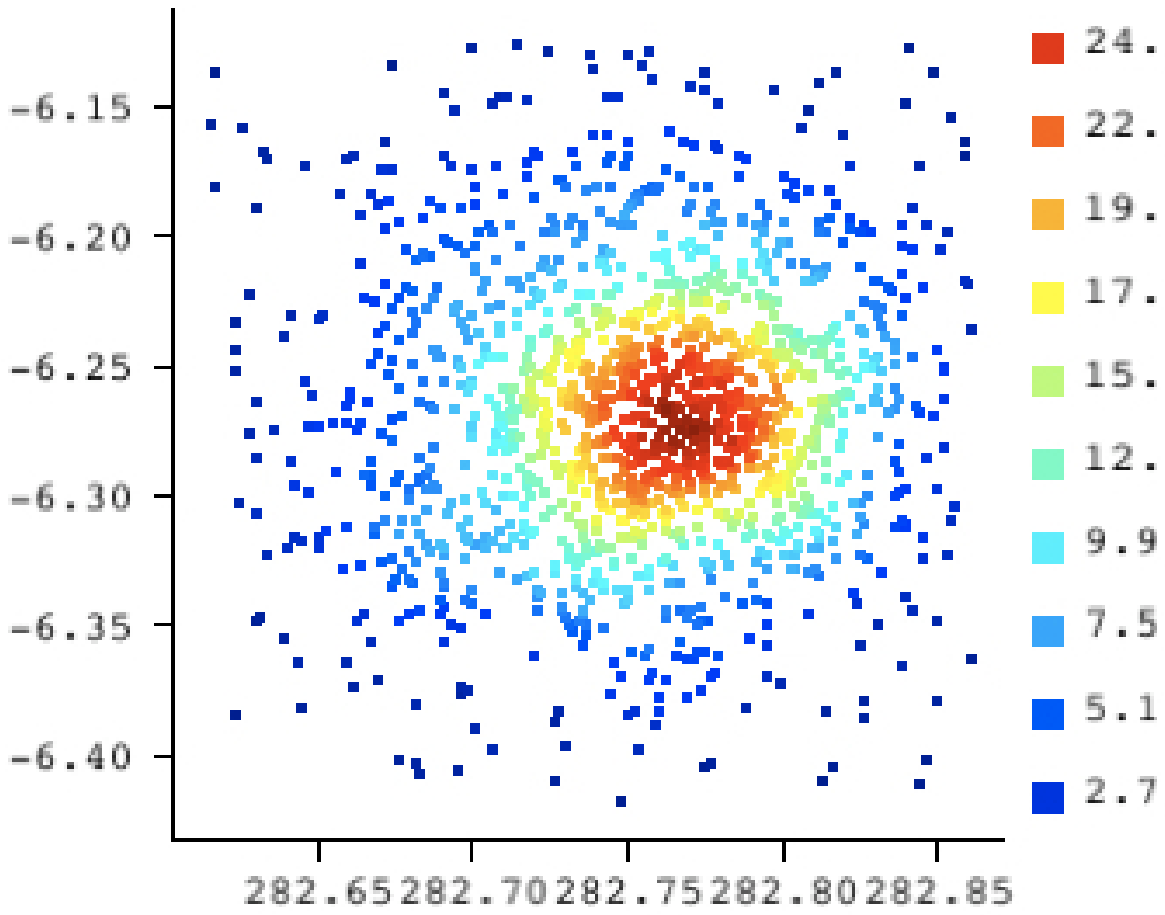}
\caption{Spatial density profiles for targets (left to right) NGC\,1981, NGC\,2516, 
NGC\,6494 and M\,11 using raw data ({\it top}) and decontaminated data ({\it down}). 
Color bars on the right represent stellar density.}
\label{f:sdp}
\end{figure}

Comparison fields were selected for the targets with the same area 
of the cluster data and similar reddening, as deduced from 
near-infrared (2MASS) and mid-infrared (IRAS) images. NGC\,1981 comparison 
field was extracted 1 deg. northwest due to the southward nebulosity associated 
with NGC\,1977. For the remaining targets circular annular fields 
centered in the cluster were used. Figure \ref{f:sdp} compares spatial density profiles 
for the targets built from raw data and the decontaminated subsample, discussed
below.

\firstsection
\section*{New Center Determination}

Precise determination of the clusters' central coordinates is an essential step 
before a reasonable radial density profile can be obtained. Catalogued center values 
are mainly intended for identification purposes and generally imprecise. Therefore, 
we have estimated the center of the targets by first selecting a region around the 
coordinates given by DAML02. 
Furthermore the selected region was divided in bins of right ascension and 
declination and star counts were made inside them. We used these star counts to 
build spatial profiles and fit a gaussian function to the star's distribution on 
RA and DEC. The center coordinates were initially taken as the center of the fitted 
gaussian.

This procedure was applied to different bin sizes and the corresponding center 
coordinates were used to create histograms showing the most recurrent 
RA and DEC. No trends were found between the coordinates and bin size and we 
adopted these most recurrent values as the  new center coordinates of the cluster. 
Table \ref{t:center} show the results of the applied method.

\begin{table}[h]
\scriptsize
\centering
\caption{({\it left}) Center determination and their difference to the cluster center 
in DAML02 catalog. ({\it right}) Determined King parameters and uncertainties.}
\begin{tabular}[h]{|l|c|c|c||c|c|c|}\hline
		  & \multicolumn{3}{|c||}{Center Determination}  &  \multicolumn{3}{c|}{King profile parameters} \\ \hline
Cluster   & $\alpha_{new}$              & $\delta_{new}$                                    & $\Delta_{lit}$ & $\sigma_{0}$     &   $R_C$         & $\sigma_{bg}$\\ \hline
NGC1981   & +05$^h$\, 35$^m$\, 07.9$^s$ & -04$^\circ$\, 20$^\prime$\, 34$.8^{\prime\prime}$ & 3.57$^\prime$  & 2.61 \,  (0.3)   &   1.0 \, (0.2)  & 0.51 \,(0.05)\\
NGC2516   & +07$^h$\, 57$^m$\, 42.0$^s$ & -60$^\circ$\, 43$^\prime$\, 22.8$^{\prime\prime}$ & 3.35$^\prime$  & 14.9 \,  (2.0)   &   0.5 \, (0.1)  & 0.95 \,(0.07)\\
NGC6494   & +17$^h$\, 56$^m$\, 37.7$^s$ & -19$^\circ$\, 00$^\prime$\, 32.4$^{\prime\prime}$ & 5.49$^\prime$  & 17.8 \,  (0.8)   &   2.9 \, (0.2)  & 10.7 \,(0.4) \\
M11       & +18$^h$\, 51$^m$\, 02.2$^s$ & -06$^\circ$\, 15$^\prime$\, 32.4$^{\prime\prime}$ & 0.71$^\prime$  & 42.9 \,  (2.5)   &   2.1 \, (0.2)  & 4.0  \,(1.3) \\
\hline
\end{tabular}
\label{t:center}
\end{table}

\firstsection \firstsection
\section*{Membership Probability Assignment}

For membership assignment we built CMDs for both cluster and field stars and divided
the diagrams into small cells in $J$ and $J-H$ axes. The cells are small 
enough to detect local variations of field-star contamination on the various 
sequences in the CMD, but large enough to accommodate a significant number of 
stars. Typical cell sizes are $\Delta J=0.5$\,mag. and $\Delta (J-H)=0.25$\,mag., with 
dense clusters accepting smaller cells while the sparse ones requiring larger ones.
Membership probabilities are assigned to cluster's stars within each cell based on 
the overdensity of cluster stars with relation to the field stars, 
according to the relation $P=(N_{clu}-N_{fld})/N_{clu}$. Null probability was 
assigned whenever an excess of field stars over cluster stars occurred in a given cell.

\begin{figure}
\centering
\includegraphics[scale=0.29,angle=0]{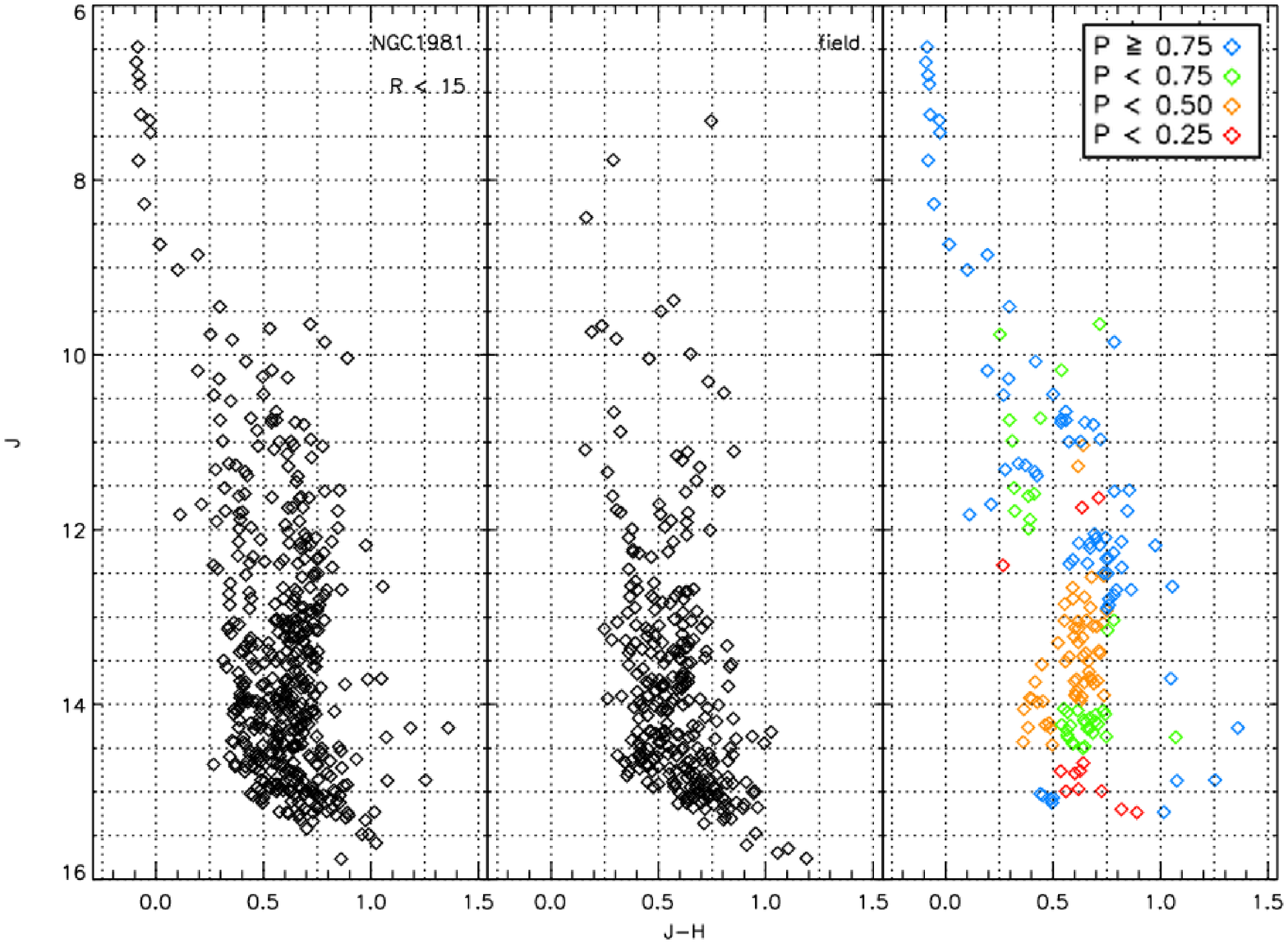}
\includegraphics[scale=0.29,angle=0]{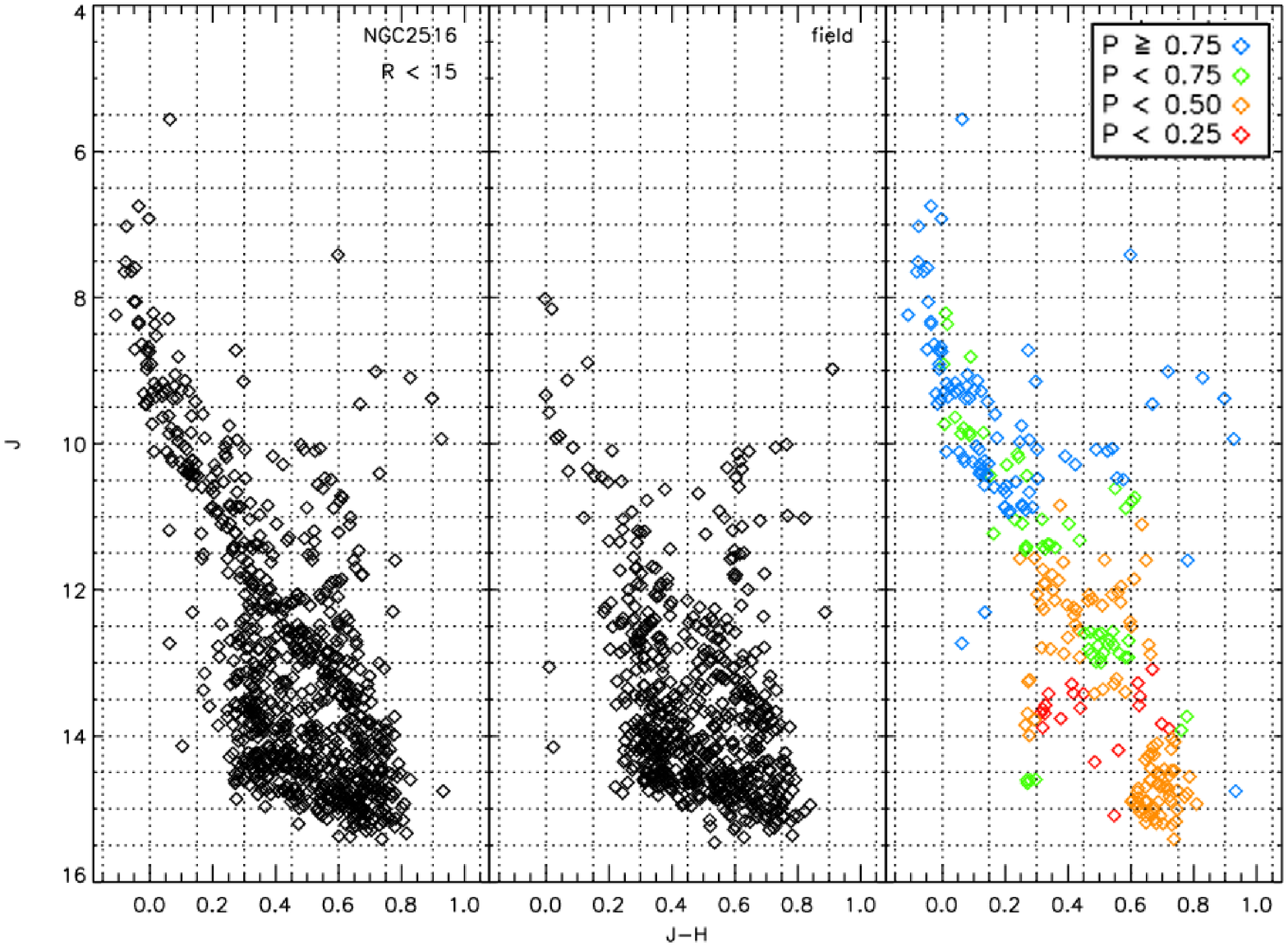}\\
\includegraphics[scale=0.29,angle=0]{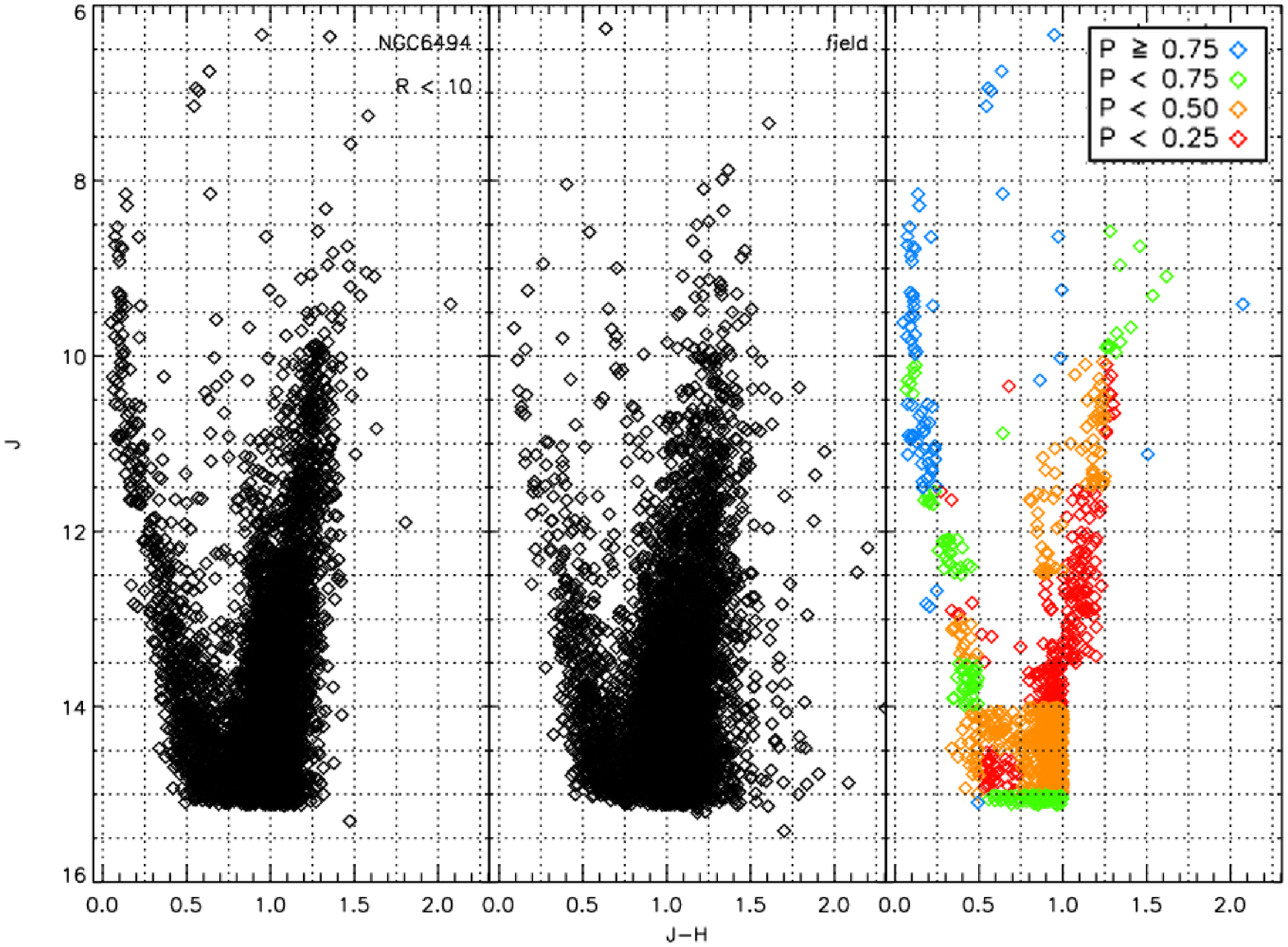}
\includegraphics[scale=0.29,angle=0]{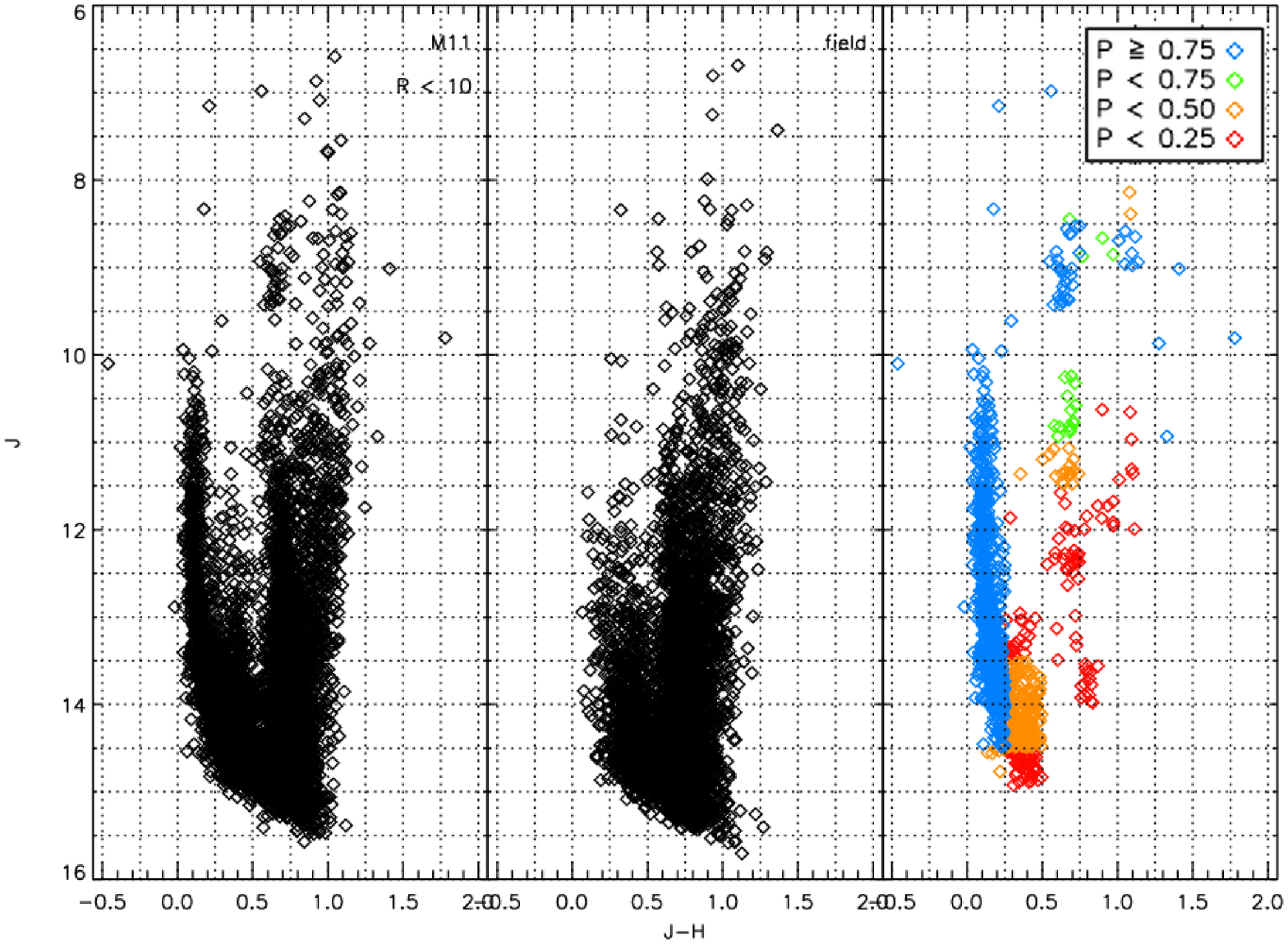}\\
\caption{Cluster and control field CMDs. From top left to bottom right: NGC1981; 
NGC 2516; NGC 6494; M11. Cluster raw data ({\it left panel}), control field 
({\it middle panel}), cluster cleaned subsample ({\it right panel}). Color-coded 
inset show the assigned membership probability.}
\label{f:cmds}
\end{figure}

For the decontamination of the CMD, a subset of the original cluster sample was  
created by removing from each cell in the cluster's CMD, the expected number of 
field stars as measured in the control field CMD. Stars where removed based on their
distance to the center of the cluster, and cells without cluster overdensity had all
stars inside their limits removed. We will refer to this final subsample as clean 
subsample hereafter. Fig. \ref{f:sdp} shows the spatial density profiles of the targets 
using the raw data and the clean subsample. Fig. \ref{f:cmds} shows the cluster, field 
and clean CMD for the target clusters.

\section*{Radial Density Profile}

\begin{figure}
\centering
\includegraphics[scale=0.25]{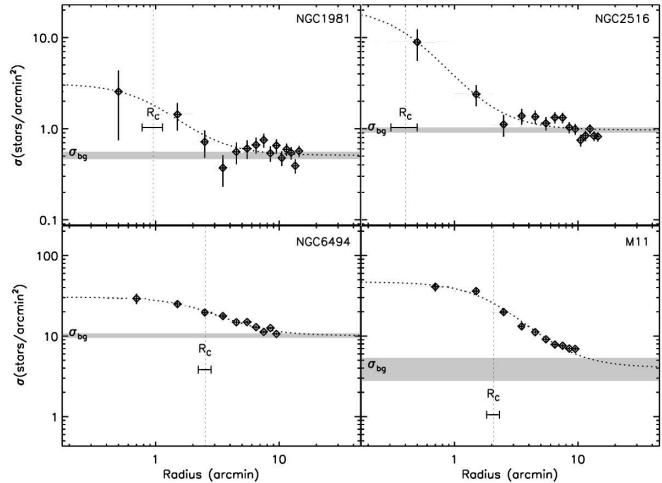}
\caption{King profile fit on the target clusters. From top left to bottom right: NGC\,1981, NGC\,2516, NGC\,6494 and M\,11. Core radius and background density are indicated.} 
\label{f:rdp}
\end{figure}

In order to probe the spatial profile and derive structural properties of the 
targets, radial density profiles (RDP) were built by counting stars inside rings of 
one arcmin width, centered in the cluster and then dividing by the area of the ring. 
King profile fitting used the modified function 
$\sigma(R) = \sigma_{bg} + \sigma_0/(1+(R/R_C)^2)$, as introduced in \citet{King}. 
Fig. \ref{f:rdp} shows RDP around the new center coordinates for the target clusters 
using the clean subsample.

Table \ref{t:center} shows the central surface density $\sigma_{0}$, core radius 
$R_C$ and background surface density $\sigma_{bg}$, with their respective 
uncertainties, for the decontaminated samples. The values 
found for M\,11 agree, within the uncertainties, to those determined by 
\citet{sbb05}.

\section*{Conclusion}

By removing stars similar to the field population in the cluster region, the method
effectively leaves the final subsample of stars less contaminated by the background.
Additionally it provides membership probabilities for all cluster stars, removing 
stars with null probability from the sample. This subsample allows for a much better
defined cluster sequences in the CMD, providing good conditions for subsequent 
isochrone fitting and more accurate values for cluster's parameters, as can be clearly
seen in Fig. \ref{f:sdp} and Fig. \ref{f:cmds}.

{
\end{document}